\documentclass[a4paper,fleqn,usenatbib]{mnras}

\usepackage{newtxtext,newtxmath}

\usepackage[T1]{fontenc}

\DeclareRobustCommand{\VAN}[3]{#2}
\let\VANthebibliography\thebibliography
\def\thebibliography{\DeclareRobustCommand{\VAN}[3]{##3}\VANthebibliography}

\usepackage{graphicx}	
\usepackage{amsmath}	
\usepackage{multicol}        
\usepackage{bm}		
\usepackage{pdflscape}	
\usepackage{natbib}

\title[]{Relationship between magnetic field properties and statistical flow using numerical simulation and magnetic feature tracking on solar photosphere
}
\author[K. Takahata et al. 2021]{K. Takahata$^{1}$\thanks{Contact e-mail: \href{taka8kemo@yahoo.co.jp}{taka8kemo@yahoo.co.jp}},
  H. Hotta$^{1}$,
  Y. Iida$^{2}$,
  T. Oba$^{3}$
\\
$^{1}$Department of Physics, Graduate School of Science, Chiba University, 1-33 Yayoi-cho, Inage-ku, Chiba 263-8522, Japan \\
$^{2}$Department of Engineering, Graduate School of Science and Technology, Niigata University, 2-8050, Ikarashi, Nishi-ku, Niigata City, Niigata 950-2181, Japan\\
$^{3}$Institute of Space and Astronautical Science (ISAS), Japan Aerospace Exploration Agency, 3-1-1 Yoshinodai, Chuo-ku, Sagamihara, Kanagawa 252-5210, Japan
}
\date{Accepted XXX. Received YYY; in original form ZZZ}

\pubyear{2021}
\begin{document}
\label{firstpage}
\pagerange{\pageref{firstpage}--\pageref{lastpage}}
\maketitle

\begin{abstract}
	We perform radiative magnetohydrodynamic calculations for the solar quiet region to investigate the dependence of statistical flow on magnetic properties and the three-dimensional (3D) structure of magnetic patches in the presence of large-scale flow that mimics differential rotation. It has been confirmed that strong magnetic field patches move faster in the longitudinal direction at the solar surface. Consequently, strong magnetic patches penetrate deeper into the solar interior. The motion of the deep-rooted magnetic patches is influenced by the faster differential rotation in the deeper layer. In this study, we perform realistic radiative magnetohydrodynamic calculations using R2D2 code to validate that stronger patches have deeper roots. We also add large-scale flow to mimic the differential rotation. The magnetic patches are automatically detected and tracked, and we evaluate the depth of 30,000 magnetic patches. The velocities of 2.9 million magnetic patches are then measured at the photosphere. We obtain the dependence of these values on the magnetic properties, such as field strength and flux. Our results confirm that strong magnetic patches tend to show deeper roots and faster movement, and we compare our results with observations using the point spread function of instruments at the Hinode and Solar Dynamics Observatory (SDO). Our result is quantitatively consistent with previous observational results of the SDO.
\end{abstract}




\section{Introduction}
The solar surface is filled with turbulent thermal convection and magnetic fields. Energy is continuously generated by nuclear fusion around the centre of the Sun. The energy is transported by the radiation in the radiation zone ($<0.71R_{\sun}$). In the outer 30\% of the solar interior, the energy is transported by thermal convection \citep{2014masu.book.....P}. Thus, the solar surface is filled with granulation, with thermal convection cells with a scale of about 1 Mm. Due to the turbulent thermal convection, magnetic fields of various scales are observed at the solar surface.
\cite{2009ApJ...698...75P} found the power-law distribution of the magnetic flux from $10^{17}$ Mx to $10^{23}$ Mx, and the power-law index is --1.85.
\par
While the solar convection zone is filled with turbulent thermal convection, solar physicists have revealed the existence of large-scale flows called differential rotation and meridional circulation.
Such large-scale flows are essential for understanding the 11-year solar cycle as magnetic flux generation and transport \citep{1955ApJ...122..293P,1999ApJ...518..508D}. Helioseismology is a great tool to evaluate the solar internal flow accurately \citep{1998ApJ...505..390S,2003ARA&A..41..599T}. The differential rotation is known to have a near-surface shear layer in which the angular velocity decreases in the radius direction near the solar surface.
\par
Another way to measure the large-scale flow in the solar interior is magnetic feature tracking. There are various sizes of magnetic patches on the solar surface. We can intuitively think that the larger and stronger patches tend to penetrate deeper and be influenced by the flow in the deeper layer. Therefore, the statistical magnetic patch properties, such as average magnetic field strength, magnetic flux and motion, may reveal the solar internal flow.
\par
An observational attempt has already been made to this end. \cite{2018ApJ...864L...5I} tracked the solar surface magnetic patches and obtained the relation between average magnetic field strength and the velocity of magnetic patches. Their result shows that stronger magnetic patches tend to move faster than the average differential rotation in their latitude. The deep solar flow may cause this relation because the angular velocity decreases in the radial direction in the near-surface layer. We cannot, however, obtain the detailed 3D structure of the turbulent magnetic field in the solar interior from observations and cannot validate our notion. \par
In this study, we carry out numerical simulations to model the photosphere and quantify the statistical properties of the magnetic patches.
The purpose of this study is to determine the effect of the large-scale flow on the motion of the magnetic patches observed at the photosphere. We aim to obtain information on the solar interior from surface observations by evaluating the relationship between the motion of each magnetic patch and properties, such as mean-field strength and magnetic flux. 
\par
Although helioseismology provides accurate measurements of the solar internal velocity field, the measurements require long observation times, making it difficult to follow the instantaneous flow field. 
If we can quantitatively associate the motion of magnetic features with the internal flow, it will be possible to estimate the instantaneous flow and be easier to better understand the solar interior. 

\section{Numerical simulation}
\subsection{Numerical setting}
In this study, we use radiation magnetohydrodynamics code
R2D2 \citep[Radiation and RSST for Deep Dynamics:][]{2019SciA....5.2307H,2020MNRAS.494.2523H,2020MNRAS.498.2925H}. The R2D2 code solves the following magnetohydrodynamic equations:
\begin{align}
\frac{\partial\rho}{\partial t}
&=
-\nabla\cdot(\rho \boldsymbol{v}),
\\
\frac{\partial }{\partial t}(\rho \boldsymbol{v})
&=
-\nabla\cdot(\rho \boldsymbol{v}\boldsymbol{v})
-\nabla p
+\rho \boldsymbol{g}
+\frac{1}{4\pi}
\left(\nabla\times\boldsymbol{B}\right)\times \boldsymbol{B},
\\
\frac{\partial \boldsymbol{B}}{\partial t}
&=
\nabla
\times\left(\boldsymbol{v}\times\boldsymbol{B}\right),
\\
\rho T \frac{\partial s}{\partial t}
&=
\rho T \left(\boldsymbol{v}\cdot\nabla\right)s + Q,
\\
p &= p\left(\rho,:s\right),
\end{align}
where $\rho$, $\boldsymbol{v}$, $p$, $T$, $\boldsymbol{g}$, $\boldsymbol{B}$, $s$ and $Q$ are the density, velocity, gas pressure, 
temperature, gravitational acceleration, magnetic field,
entropy and radiative heating, respectively. 
\par
The R2D2 code solves the equations with the fourth-order spatial derivative and four-step Runge--Kutta method
for time integration. The gas pressure $p$ is obtained from the entropy and the density table served by the OPAL  equation of state considering partial ionisation \citep{1996ApJ...456..902R}. 
The radiative heating $Q$ is calculated by solving the radiation transfer equation using the grey approximation and the short characteristic in two directions. 
For details, see equations (6)-(11) for $Q_{\mathrm{rad}}$ in \citet{2019SciA....5.2307H}.
For the initial stratification, we use Model S \citep{1996Sci...272.1286C}, which is the standard model of the solar interior. We use the periodic boundary condition for the horizontal directions and make the magnetic field vertical at the upper and lower boundaries. The initial value of the magnetic field is a weak and uniform magnetic field of $B_y=0.1\ \mathrm{G}$.
\par
The computational domain extends 98.304 Mm in the horizontal directions  and 24.576 Mm in the vertical direction. The number of grids is 1024 $\times$ 1024 
in the horizontal directions and 512 in the vertical direction with a grid spacing of 96 km in the horizontal directions and 48 km in the vertical direction.
This is a acceptable grid spacing to resolve the features at the phtosphere.
The upper boundary is 700 km above the average $\tau=1$ surface, where $\tau$ is the optical depth. Our setup allows us to include some supergranulations with a scale of about 30 Mm.
\par
After four days of evolution, turbulent thermal convection creates magnetic fields through small-scale dynamo \citep[e.g.,][]{2014ApJ...789..132R} observed at the photosphere. After the convection and magnetic field become a statistically steady state, we add eq. (\ref{eq:forceflow}) to the equation of motion to enforce the large-scale flow.\\
\begin{equation}
    \frac{\partial v_y}{\partial t} = - \left( \frac{\overline{v}_y (z) - v_\mathrm{f} (z)} {\tau_\mathrm{f}}  \right)
    \label{eq:forceflow}
\end{equation}
Here, $z$ is the vertical direction and $\overline{v}_y$ is the horizontally averaged $y$-component velocity. We set the reference flow $v_\mathrm{f}$ as $v_\mathrm{f}(z) = v_0z/z_\mathrm{min}$, where $v_0=500\ \mathrm{m\ s^{-1}}$. This is much faster than the real solar value ($\sim 15\ \mathrm{m\ s^{-1}}$). When we adopt the realistic value, we need to detect a massive number of magnetic patches to evaluate the slow large-scale flow against the fast turbulent convection ($\sim4\ \mathrm{km\ s^{-1}}$). Because the possible number of the detection is restricted by numerical resources, we increase the reference flow speed to decrease the required numerical cost. Because $z=0$ is the solar surface, the reference flow is $0\ \mathrm{m\ s^{-1}}$ and $500\ \mathrm{m\ s^{-1}}$ at the surface and the bottom of the computational domain, respectively. We adopt the relaxation time $\tau_\mathrm{f} = H_p/v_\mathrm{c}$, where $H_p$ and $v_\mathrm{c}$ are the pressure scale height and the typical convection velocity, respectively. The convection velocity $v_\mathrm{c}$ is evaluated with the solar energy flux at the surface $F_\odot=6.3\times 10^{10}\ \mathrm{erg\ s^{-1}\ cm^2}$ and the density $\rho_0$ as $v_\mathrm{c}=F_\odot/(2\rho_0^{1/3})$. In our setting, the large-scale flow in the deeper region is faster to mimic the near-surface differential rotation.
After the convection reaches a steady state, we set the output cadence to 45 seconds and continue the calculation for about 45 hours to obtain data for tracking.

\subsection{Point spread function}
The simulated magnetic field data are preprocessed to be compared with observations. The magnetic field data at $\tau=0.01$ surface are convolved with satellite point spread functions (PSF) to prepare three kinds of data:
(a) The raw simulated magnetic field data;
(b) Magnetic field data convolved with the PSF for the Solar Optical Telescope \citep[SOT,][]{2008SoPh..249..167T} of Hinode \citep{2007SoPh..243....3K}; and
(c) Magnetic field data convolved with the PSF for the Helioseismic and Magnetic Imager \citep[HMI,][]{2012SoPh..275..229S}
of the Solar Dynamics Observatory \citep[SDO,][]{2012SoPh..275....3P}.
The PSF for the Hinode/SOT is obtained from \citet{2009A&A...501L..19M} as follows:
\begin{equation}
    \mathrm{PSF}_{\mathrm{Hinode}}(r) =
    \sum_{i=1}^4 w_i c_ie^{-r^2/(2b_i^2)},
    \label{eq:Hinode}
\end{equation}
where the normalised constant $c_i = 1/\sqrt{2\pi b_i^2}$, weights $w_i$ and width $b_i$ were used 
with values of 5500 \AA. This wavelength is close to the commonly used Na D 589.6 nm line for Hinode's magnetogram.

The PSF for SDO/HMI is adopted from \cite{2012SoPh..275..261W},
\begin{equation}
    \mathrm{PSF}_{\mathrm{SDO}}(r) = e^{- \left( \frac{r} {w} \right)^2},
    \label{eq:SDO}
\end{equation}
where $r$ is the distance from the target grid and 
$w=0.909\ \mathrm{arcsec}$.

For a simple approach to mimic observations, the magnetic field data are directly convolved with the PSF, while the circular polarimetric intensity (Stokes $V$) should be convolved.
In order to convolute Stokes $V$, we need to calculate the polarization of the radiation. Since this process is not implemented in the R2D2 code, we just spatially convolved the magnetic field obtained in the calculations.
\subsection{Detection and tracking of magnetic features}
\subsubsection{Detection of magnetic patches}
\label{sec:Detection}
We automatically detect the magnetic patches from the simulated magnetic field data at the photosphere and evaluate the depth and the velocity of these features \citep[see][for details]{2012ApJ...752..149I}. A magnetic patch is defined as a group of numerical grids that have a magnetic field strength larger than a specific threshold value in a horizontal plane. The thresholds for magnetic field strength, area and flux at the photosphere are set to 40 G, 20 grids and $10^{17}$ Mx, respectively. In the solar interior, we only set a threshold for the magnetic field strength. Because the magnetic field strength increases with depth, we set the field strength threshold at four times the horizontally averaged field strength $\overline{|B_z|}$. We labelled each magnetic field patch that satisfies these thresholds using OpenCV. Note that if the positive and negative magnetic fields are not handled separately, neighbouring magnetic fields with the opposite sign would be considered as one feature.

\subsubsection{Depth evaluation}
\begin{figure}
    \centering
    \includegraphics[width=0.8\columnwidth]{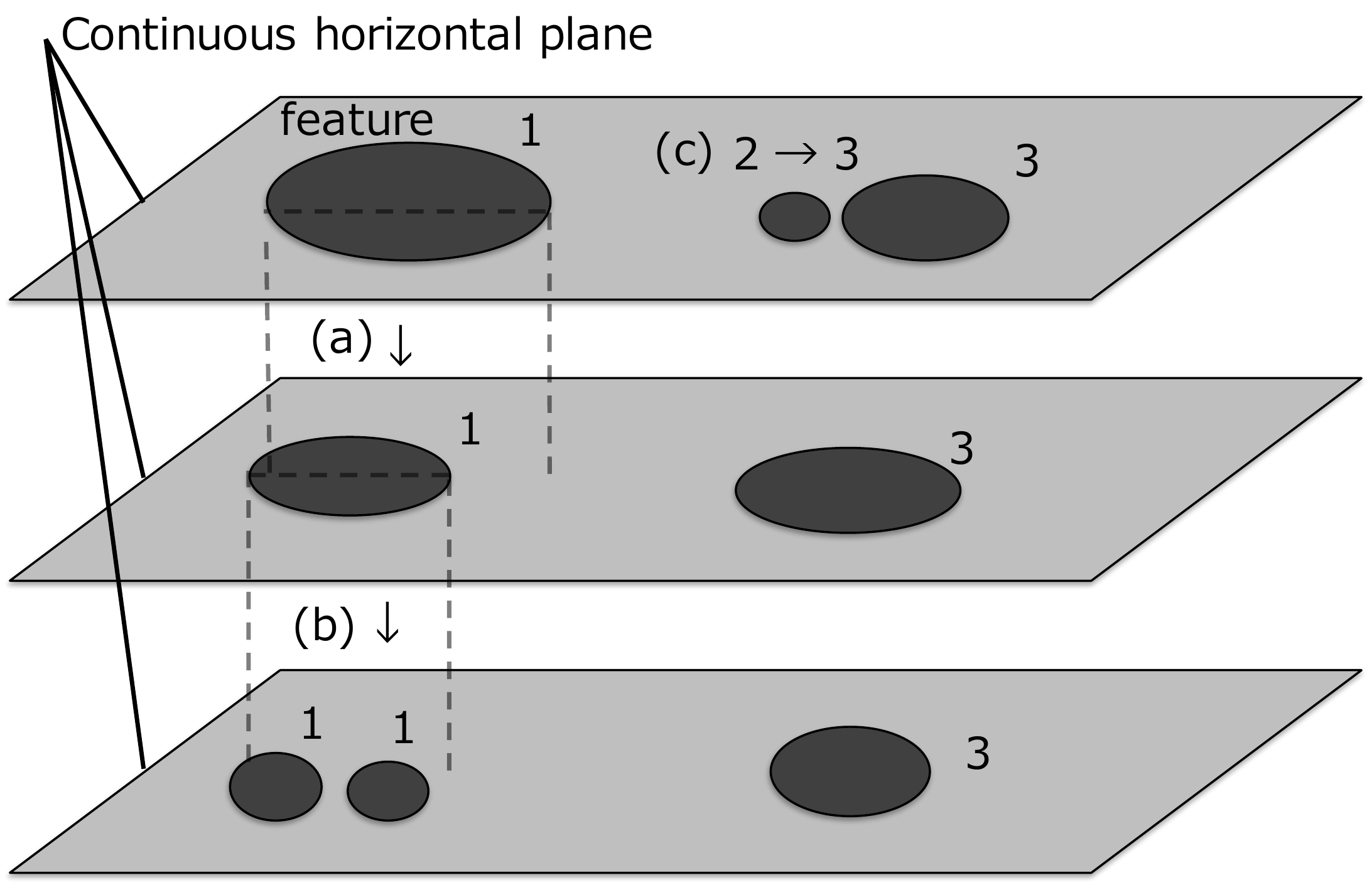}
    \caption{Schematic picture for explaining our method of evaluating the depth of the magnetic structure.
    }
    \label{fig:deepimage}
\end{figure}

In this subsection, we explain our method of evaluating the depth of magnetic patches. We expect that the magnetic structures penetrate to a deep layer and lose their coherence in a certain layer. We evaluate this tendency with automatic detection. The detected magnetic patch in \ref{sec:Detection} is associated with magnetic structures in deep layers. We compare two-slice data at $z=z_k$ and $z_{k-1}$, where $k$ denotes the vertical grid. If magnetic patches in two layers overlap, we label the patch at the plane $z=z_{k-1}$ with the same number as that at $z=z_k$ (see Fig. \ref{fig:deepimage}a).
If more than one patch in the plane $z=z_k$ overlaps with a patch at $z=z_{k-1}$, we adopt the label number of the largest patch at $z=z_k$ for the overlapped patch at $z=z_{k-1}$. The same label number is assigned to all magnetic patches at every height for a magnetic structure, even if they are split (see Fig. \ref{fig:deepimage}b). If several patches are merged, the label number of the largest overlapped patch is used for the other merged features (Fig. \ref{fig:deepimage}c). These procedures are carried out for 505 planes from the plane near the photosphere to the bottom of the computational domain, and the amount of magnetic flux of each label and the deepest position reached are recorded. 
This processing series was performed for 34 steps of data with a cadence of 30 minutes.
The structure of about 30,000 magnetic patches was then analysed.


\subsubsection{Tracking of the horizontal direction at the photosphere}
\begin{figure}
    \begin{center}      
    \includegraphics[width=0.55\columnwidth]{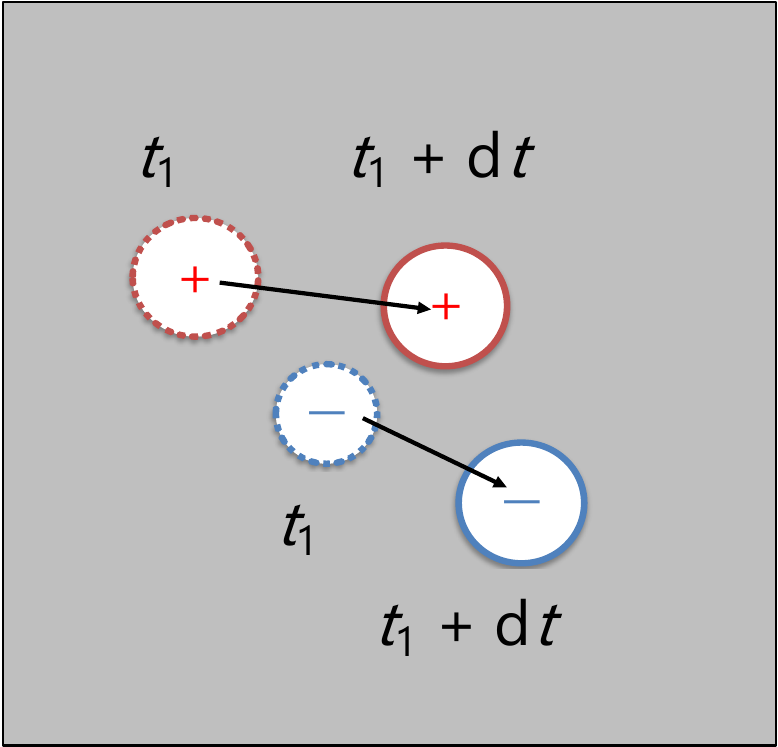}
    \caption{Schematic for detecting the velocity of the magnetic patch.
    }
    \label{fig:speedimage}
    \end{center}    
\end{figure}

In this subsection, we explain our method of measuring the velocity of the magnetic patch.
The centre of gravity weighted by vertical magnetic field strength is obtained for each magnetic patch.
\begin{align}
\boldsymbol{r}_\mathrm{g} =\frac{\int \boldsymbol{r} B_z dS}{\int B_z dS},
\label{eq:center}
\end{align}
$\boldsymbol{r}$ is the horizontal position, and the integration is carried out in a patch. To identify the patch movement, we need to find a pair of patches in two consecutive time steps. 
We define the pair as two same-signed magnetic patches with the shortest distance in two time steps. These patches are assumed to be the same, and the motion of the patches causes the distance. The velocity of the magnetic features was calculated with the movement of the centre of gravity by dividing the distance by the time cadence (see Fig. \ref{fig:speedimage}).
We have traced about 2.9 million, 1.3 million, and 330,000 magnetic patches from the original data, the Hinode PSF convolution data, and the SDO PSF convolution data, respectively.
\section{Results}

\begin{figure}
    \centering
    \includegraphics[width=0.75\columnwidth]{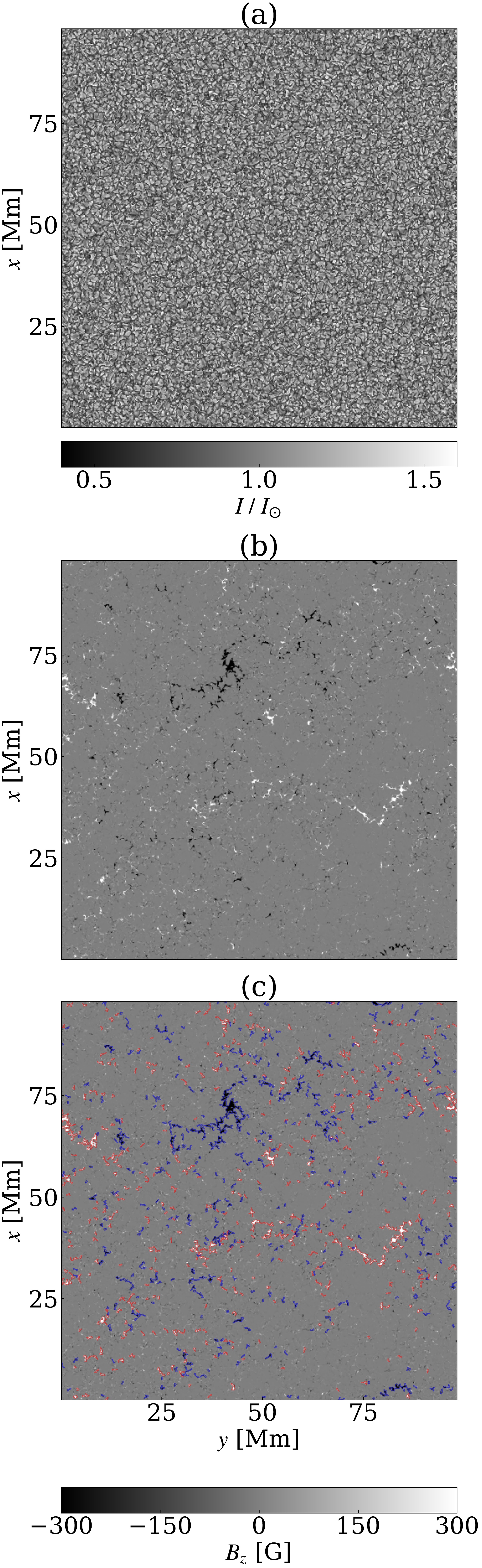}
    \caption{The radiative intensity (panel a), the vertical magnetic field strength (panel b) and the detection of magnetic patches (panel c). The intensity is normalised by the average intensity $I_{\sun}$. Panels b and c show the vertical magnetic field at $\tau=0.01$ surface. The red and blue lines indicate positive and negative magnetic patches, respectively.
    }
    \label{fig:photo1}
    
\end{figure}

Fig. \ref{fig:photo1} shows the raw simulation data comprising the radiative intensity (panel a), the vertical magnetic field strength at $\tau=0.01$ (panel b) and the magnetic field strength and detected magnetic patches with red and blue lines (panel c). The intensity is normalised with the average intensity. The white and black magnetic fields indicate the positive and negative magnetic fields, respectively. The magnetic field is distributed in a network with a scale of several 10 Mm, and the structure of supergranulation can be seen. We note that although the maximum value of the colour bar for the magnetic field is 300 G, the maximum magnetic field strength exceeds 1000 G. 
The red and blue contours in panel c indicate positive and negative magnetic patches, respectively.

\begin{figure}
    \centering
    \includegraphics[width=0.8\columnwidth]{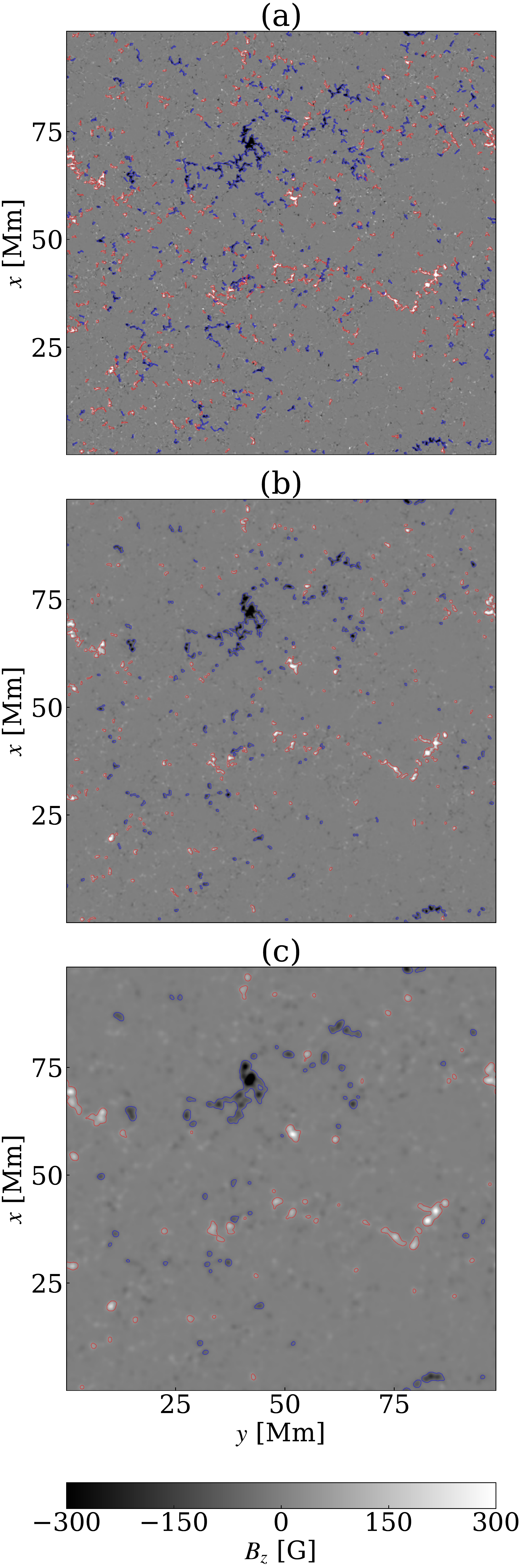}
    \caption{Magnetic field strength convolved with different kinds of PSF. The simulated raw magnetic field data (panel a), data convolved with the PSF of Hinode/SOT PSF (panel b) and data convolved with the PSF of SDO/HMI (panel c).
    }
       \label{fig:photo2}
\end{figure}

Fig. \ref{fig:photo2} shows the vertical magnetic field convolved with PSF and the detected features. 
The figure shows the raw simulated magnetic field (panel a), the magnetic field with the PSF of the Hinode/SOT (panel b) and the magnetic field with the PSF of SDO/HMI (panel c). SDO/HMI (panel c) shows the most diffused patches, and the field strength is weak on average.

\begin{figure}

    \includegraphics[width=0.8\columnwidth]{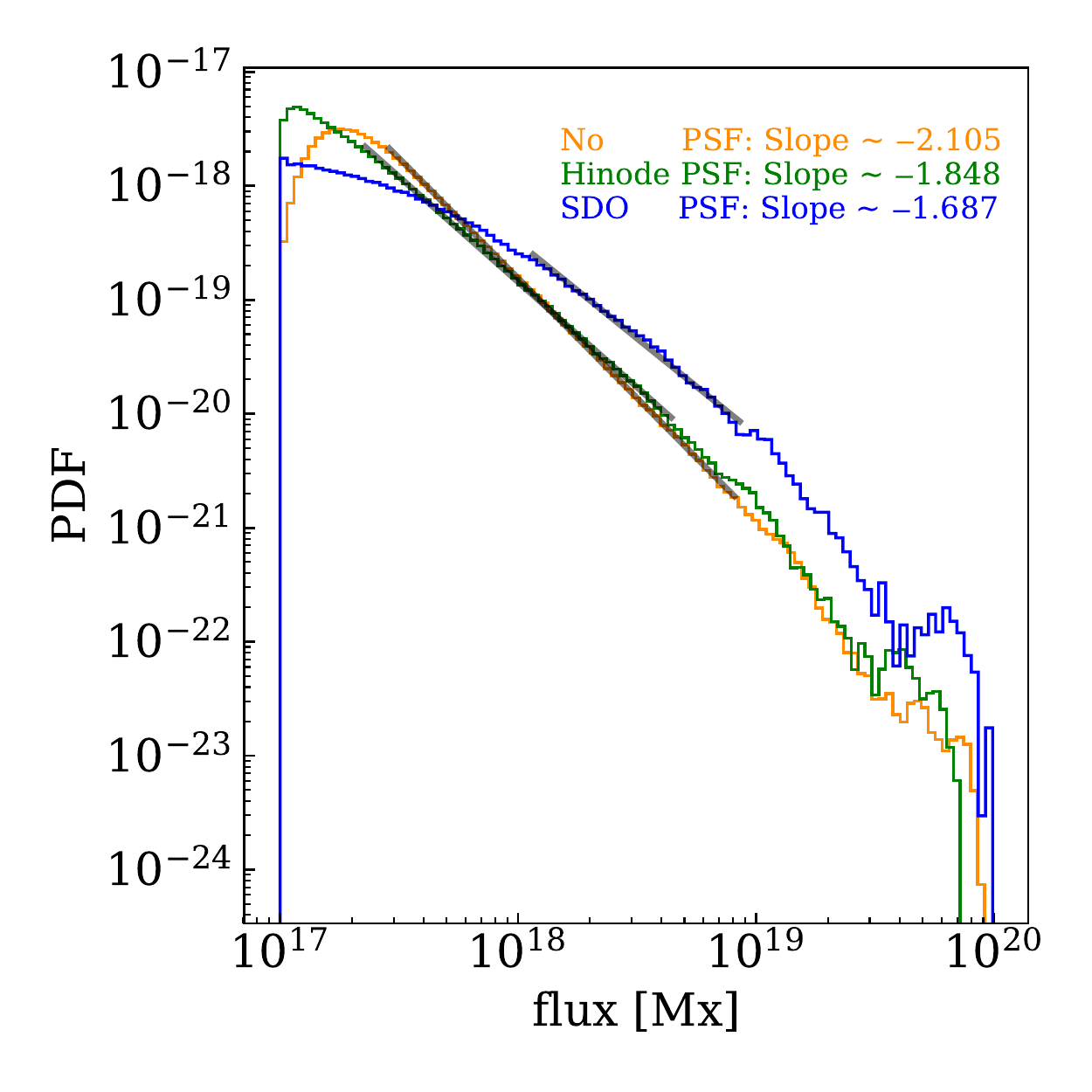}
    \caption{The PDF at the photosphere.
    Orange, green and blue lines indicate the PDF of the raw simulation data, the data with Hinode/SOT's PSF and the data with SDO/HMI's PSF, respectively. The black lines are the fitted lines for their power-law distribution.
    }
    \label{fig:pdf}

\end{figure}

Fig. \ref{fig:pdf} shows the probability density function (PDF) of the magnetic flux at the photosphere.
Orange, green and blue lines indicate the magnetic flux in the original raw data, the data with the PSF of Hinode/SOT and the data with the PSF of SDO/HMI, respectively. The black line indicates the fitting result for a range that clearly shows the power-law distribution. The slope increases with increasing resolution, i.e. the raw simulation data show a steeper slope than the data with Hinode and SDO PSF. The power-law index of the PDF with Hinode's PSF (green line) is --1.848, consistent with \cite{2009ApJ...698...75P}, which showed --1.85 for the power-law index with the Hinode observation. 
This suggests that the simulated magnetic field data reproduce the real Sun nicely 
and supports the appropriateness of the model in this study.

\begin{figure}
    \includegraphics[width=\columnwidth]{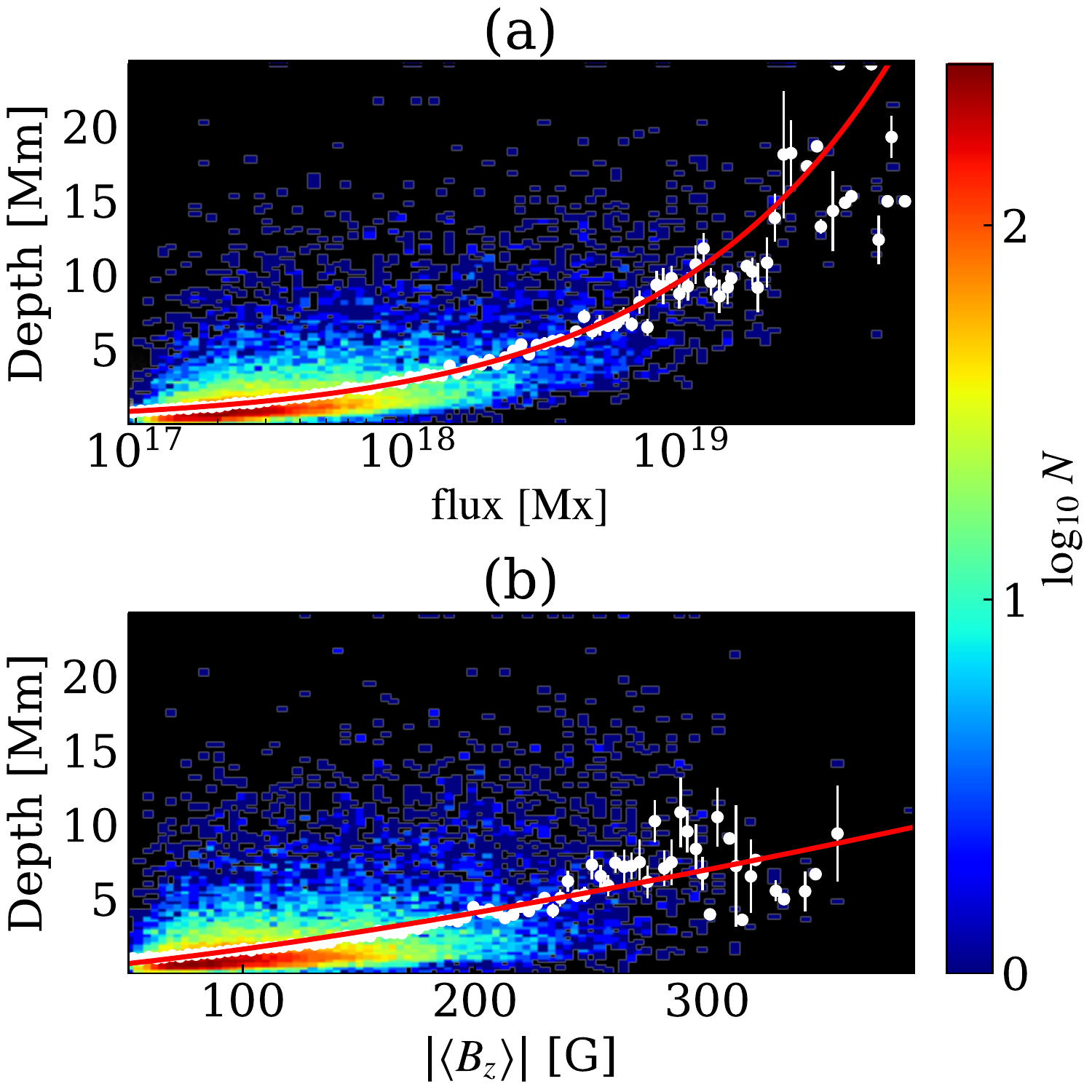}
    \caption{Detected depth of the magnetic field. The horizontal axes are the magnetic flux (panel a) and the vertical magnetic field strength (panel b). The red lines are the fitting results. The background 2D histograms are the distributions of the sample number $N$.
    }
       \label{fig:deepresult}
\end{figure}

Fig. \ref{fig:deepresult} shows the results of our analysis for the detected magnetic field depths.
$|\langle B_z\rangle|$ means the absolute value of the vertical magnetic field averaged over each magnetic patches. We divide the horizontal axis into 100 bins. Each bin has the same size in logarithmic and linear scales for magnetic flux (panel a) and magnetic field strength (panel b), respectively. Error bars indicate standard errors. The red lines indicate the fitting results. We use a power-law function to fit the result. The background two-dimensional (2D) histogram shows the distribution of the sample number $N$. The reliability decreases around $10^{20}$ Mx and 300 G because a small number of data exists and the error becomes large. This figure shows that magnetic patches with strong (weak) magnetic field strength and flux tend to have a deep (shallow) root.

\begin{figure}
    \includegraphics[width=\columnwidth]{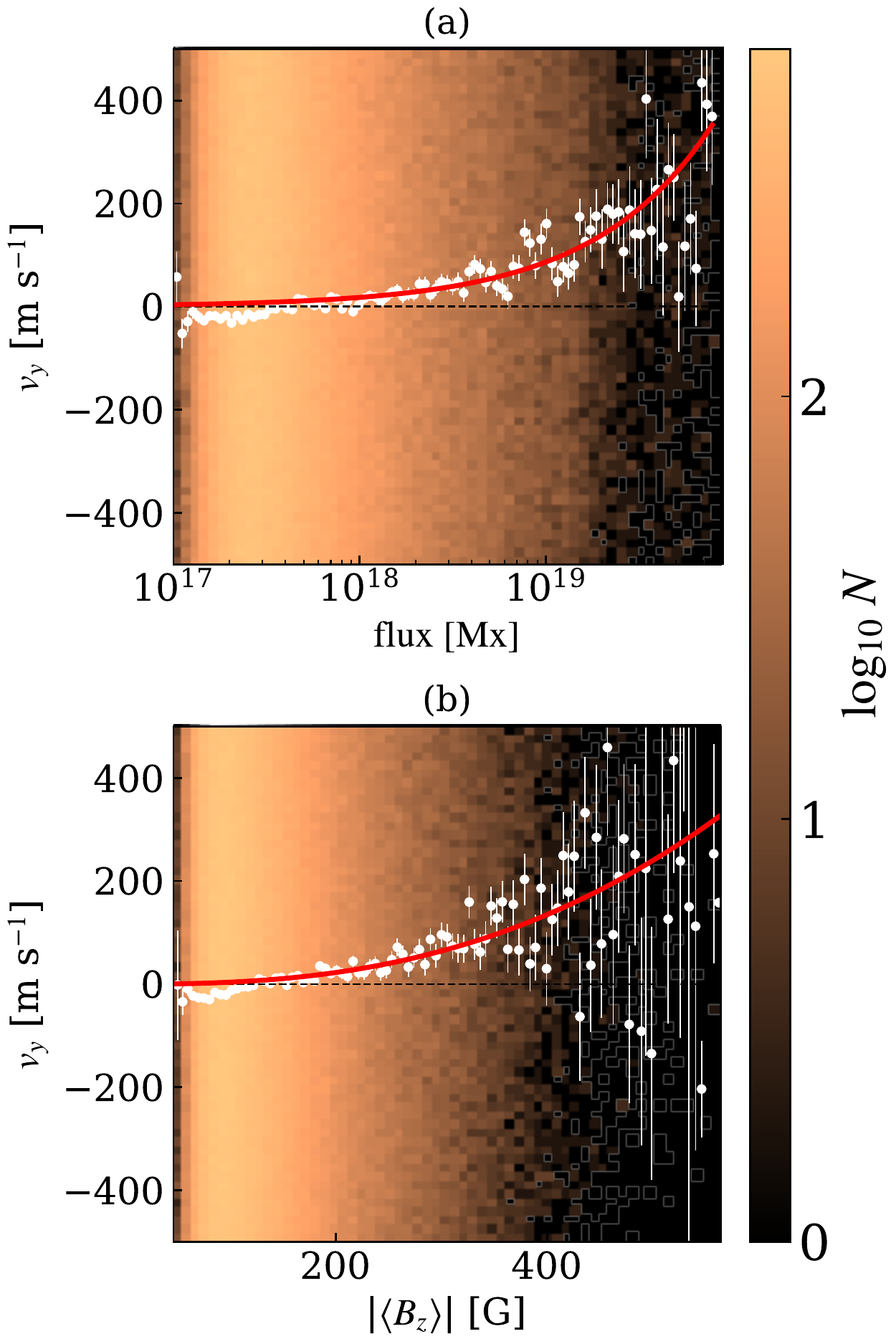}
    \caption{The velocity of the detected magnetic field at the photosphere. 
    The horizontal axes are the same as those in Fig. \ref{fig:deepresult}.
    The red lines are the fitting results. The background 2D histograms show the distribution of the number of data $N$.
    }
       \label{fig:speedresult}
\end{figure}

Fig. \ref{fig:speedresult} shows the result of the relationship between the velocity and the magnetic patch properties.
We adopt the same method as Fig. \ref{fig:deepresult}. We divide the horizontal axis into 100 bins, and the velocity is averaged over a bin. The fitting is done with a power-law function for both results. Our results show that stronger and larger magnetic patches tend to show faster movement in the large-scale flow direction.

Table \ref{tab:coff} shows coefficients of the power-law function for the fitting. The depth of the magnetic structure $d_\mathrm{m}$ and the movement velocity $v_y$ is approximated with two functions as:
\begin{align}
    & 10^\mathrm{-a}\times \Phi^\mathrm{b}, \\
    & 10^\mathrm{-c}\times\langle B_z\rangle^\mathrm{d},
\end{align}
where $\Phi$ is the magnetic flux. The fitting results for the depth $d_\mathrm{m}$ and movement velocity $v_y$ are seen in Figs. \ref{fig:deepresult} and \ref{fig:speedresult} as red lines. These results are from the raw simulation data, but the Hinode/SOT and SDO/HMI data show similar values.

\begin{table}
    \centering
    \caption{Coefficients of the power-law function for the fitting results of the magnetic structure depth.
    }
     \begin{tabular}{|l|c|c|c|c|c|} 
      \hline
      Coefficient of power-law & a & b & c & d  \\ 
      \hline

      Depth \hspace{1.25cm}[Mm] & 8.97 & 0.524 & 2.44 & 1.32 \\
      $v_y$ (No\hspace{5.5mm}PSF) [$\mathrm{m\ s^{-1}}$] & 11.0 & 0.682 & 4.58 & 2.58 \\
      $v_y$ (Hinode PSF) [$\mathrm{m\ s^{-1}}$] & 9.15 & 0.589 & 2.78 & 2.10 \\
      $v_y$ (SDO\hspace{3.5mm}PSF) [$\mathrm{m\ s^{-1}}$] & 8.18 & 0.537 & 0.28 & 1.10 \\
      
      \hline
     \end{tabular}
     
     \label{tab:coff}
  \end{table}

\begin{figure}
    \includegraphics[width=\columnwidth]{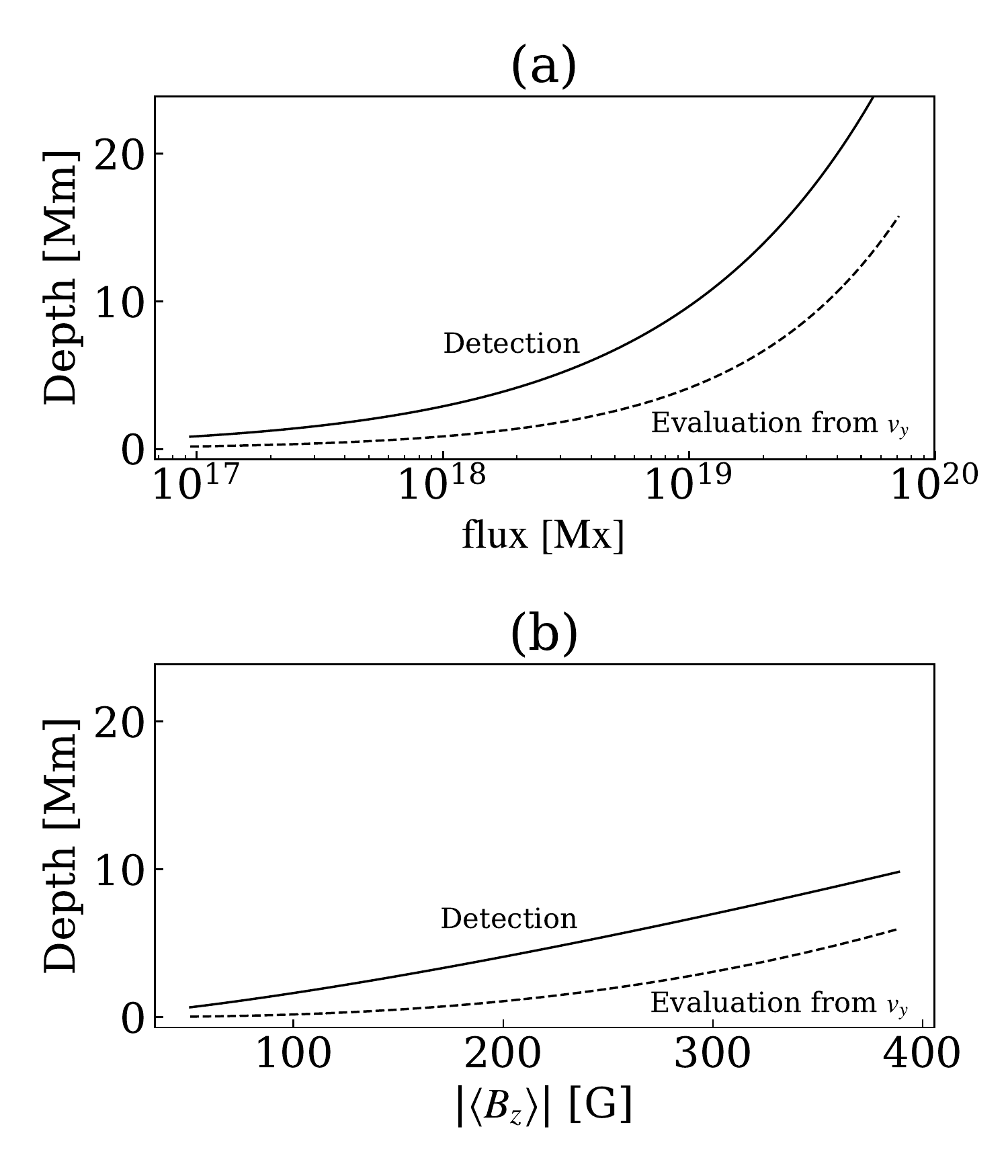}
    \caption{Comparison of the depth of the magnetic structure between direct detection (solid line) and the conversion with eq. (\ref{eq:conversion}), (dashed line).
    }
       \label{fig:all}
\end{figure}

In this study, we know the profile of the large-scale flow. The flow speed at the photosphere may be related to the large-scale flow speed in a certain depth. Because the reference flow is expressed as $v_\mathrm{f}=v_0z/z_\mathrm{min}$, the detected velocity $v_y$ at the photosphere can be converted to depth as:
\begin{align}
    z = z_\mathrm{min}v_y/v_0 \label{eq:conversion}.
\end{align}
This evaluation would be different from the direct detection of the magnetic structure depth. Fig. \ref{fig:all} shows a comparison of the depth between the direct detection and the evaluation from the detected velocity $v_y$. 
The solid line is the fitting function of the depth in Fig. \ref{fig:deepresult}. The dashed line is obtained from the conversion of eq. (\ref{eq:conversion}).
In all ranges, the solid line shows a deeper layer than does the dashed line.
\section{Discussion and Conclusion
}
In this paper, we perform numerical simulations with large-scale forced flow mimicking the solar differential rotation. By using magnetic feature tracking, we investigate the dependence of the depth of the magnetic structure and the velocity on magnetic properties (strength and flux) at the photosphere. Our results show that stronger and larger magnetic patches tend to show deeper roots and faster movement influenced by the large-scale flow. Our result is consistent with \cite{2018ApJ...864L...5I}, where the stronger magnetic field moves faster than the average differential rotation.
\par
The results in \cite{2018ApJ...864L...5I} using SDO/HMI data show a linear relationship between magnetic field strength ($>100\ \mathrm{G}$) and the movement velocity. Our result with the PSF of SDO/HMI is consistent with \cite{2018ApJ...864L...5I}. We use the power-law function for the fitting, and the power-law index is 1.10 for the relationship between the magnetic field strength and the velocity (bottom of Table \ref{tab:coff}). This means that our result can be also approximated with a linear function.
\par
In this study, we use the reference flow of 500 $\mathrm{m\ s^{-1}}$, which is much faster than the expected value in the sun. The depth of our calculation box is 25 Mm, and the expected relative velocity by the differential rotation is about 15 $\mathrm{m\ s^{-1}}$, as evaluated from \citep{1996Sci...272.1300T,2000Sci...287.2456H}. 
If we change the maximum value of the flow, the result also changes.
The turbulence convection velocity at the photosphere is 4 $\mathrm{km\ s^{-1}}$, and it is difficult to determine the tiny large-scale flow of 15 $\mathrm{m\ s^{-1}}$. Thus, we decided to adopt the fast reference velocity. To determine the real tiny flow, we need to calculate for a long period and/or prepare a large computational box to increase the number of samples. The other possible way of obtaining the result with the realistic slow large-scale is to investigate with different reference velocities. In this paper, we choose a value of 500 $\mathrm{m\ s^{-1}}$ for the reference large-scale flow to clarify the effect of the deep differential rotation. When we repeat the investigation with different reference velocities, we can extrapolate the result to the solar value in the future.
\par
\begin{figure}
  \includegraphics[width=\columnwidth]{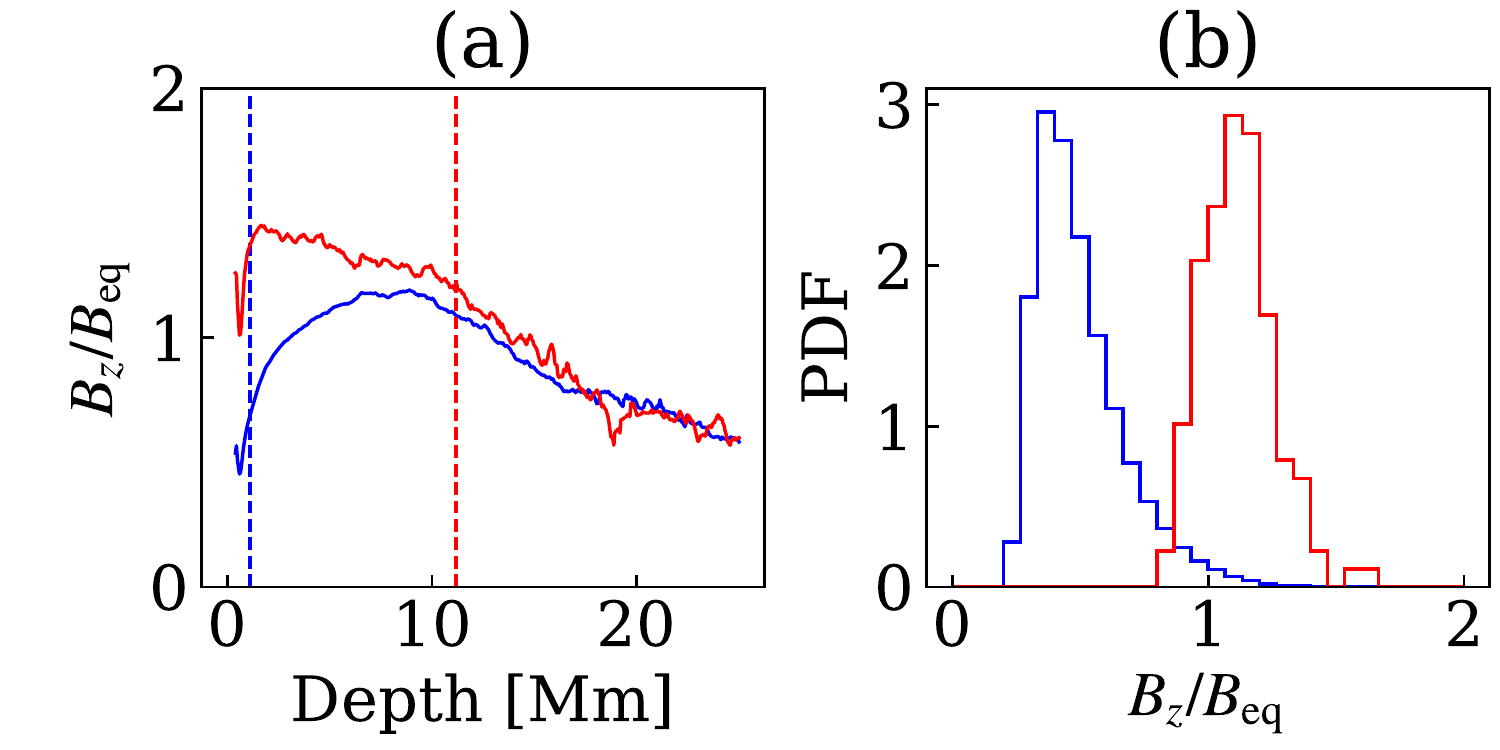}
  \caption{Panel a shows the average of $B_z/B_\mathrm{eq}$ at different depths. 
  The red and blue lines shows groups with the magnetic flux at the surface of $\Phi>3\times10^{18}$ and $\Phi<3 \times 10^{18}$ Mx, respectively.
  The dashed line shows the depth evaluated from $v_y$ (The dashed line of Fig.\ref{fig:all}) averaged over each group.
  Panel b shows the probability density function of $B_z/B_\mathrm{eq}$ for each group at the solar surface.}
  \label{fig:Beq}
\end{figure}

Fig. \ref{fig:all} shows that the depth evaluated from the detected velocity is shallower than the direct detection. 
This can be explained with the balance between kinetic and magnetic energies.
Fig. \ref{fig:Beq}a shows the average of $B_z/B_\mathrm{eq}$ at different depths, where $B_z$ is the average magnetic field strength in the magnetic patches and $B_\mathrm{eq}$ is the equipartition magnetic field strength, i.e. ${B_{\mathrm{eq}}^2}/(8\pi) =\rho (v_x^2 + v_y^2)/2$, averaged over the horizontal computational domain. 
The red and blue lines show groups with the magnetic flux at the surface of $\Phi>3\times10^{18}$ and $\Phi<3 \times 10^{18}$ Mx, respectively.
The dashed line shows the depth evaluated from $v_y$ (dashed line of Fig.\ref{fig:all}) averaged over each group.
Panel b shows the probability density function of $B_z/B_\mathrm{eq}$ for each group at the solar surface.
When the magnetic energy is smaller than the kinetic energy, i.e., $B_z/B_\mathrm{eq}<1$ at the photosphere (blue line), turbulence is already dominant near the surface. Thus, the $v_y$ of the magnetic patches at the surface is random, and the depth evaluated from $v_y$ is nearly zero.
When $B_z/B_\mathrm{eq}>1$ at the photosphere (red line), the magnetic energy dominates near the surface, and the turbulent energy dominates as the depth increases. The depth evaluated from $v_y$ (red dashed line) correspoinds to a depth where $B_z/B_\mathrm{eq}\sim1$, suggesting that 
magnetic fields with sufficient flux at the surface reflect the information at the depth of $B_z/B_\mathrm{eq}\sim1$.

\section*{Acknowledgements}
The results were obtained using Fujitsu PRIMERGY CX600M1/CX1640M1 (Oakforest-PACS) at the Joint Center for Advanced High Performance Computing (JCAHPC; proposal Nos. are hp190183 and hp200124) and Cray XC50 at the Center for Computational Astrophysics, National Astronomical Observatory of Japan. This work was supported by MEXT/JSPS KAKENHI (grant No. JP20K14510 [PI: H. Hotta]) and MEXT as Program for Promoting Researches on the Supercomputer Fugaku (Toward a unified view of the universe: from large-scale structures to planets, grant No. J20HJ00016 [PI: J. Makino]).

\section*{Data Availability}

The simulation data and source code of the magnetic feature tracking underlying this article will be shared on reasonable request to the corresponding author.




\bsp    
\label{lastpage}
\end{document}